\documentclass[pdflatex,sn-mathphys-ay]{sn-jnl}

\usepackage{footnote}
\usepackage{soul}
\usepackage{lipsum}
\usepackage{xcolor}
\usepackage{caption}
\usepackage{subcaption}
\usepackage{float}
\usepackage{csquotes}
\usepackage{url}
\NewDocumentEnvironment{alignb}{b}{%
  \begin{align*}
  \refstepcounter{equation} #1 \tag{\theequation}
  \end{align*}
}{}
\usepackage[T1]{fontenc}
\usepackage{graphicx}%
\usepackage{multirow}%
\usepackage{amsmath,amssymb,amsfonts}%
\usepackage{amsthm}%
\usepackage{mathrsfs}%
\usepackage[title]{appendix}%
\usepackage{xcolor}%
\usepackage{textcomp}%
\usepackage{manyfoot}%
\usepackage{booktabs}%
\usepackage{algorithm}%
\usepackage{algorithmicx}%
\usepackage{algpseudocode}%
\usepackage{listings}%
\usepackage{aas_macros}

\theoremstyle{thmstyleone}%
\theoremstyle{thmstyletwo}%

\theoremstyle{thmstylethree}%

\begin{document}

\title[Article Title]{Time-lag in hadronic channel: to explore delayed VHE-flare of 3C 279}


\author[1]{\fnm{Sunanda} }\email{sunanda@iitj.ac.in}

\author[2]{\fnm{Reetanjali} \sur{Moharana}}\email{reetanjali@iitj.ac.in}

\affil*[1,2]{\orgdiv{Department of Physics}, \orgname{Indian Institute of Technology}, \orgaddress{\street{Karwar}, \city{Jodhpur}, \postcode{342037}, \state{Rajasthan}, \country{Country}}}


\abstract{On 28 January 2018, the High Energy Stereoscopic System (H.E.S.S.) reported a significant very-high-energy (VHE) gamma-ray activity, occurring nearly 11 days after the high-energy (HE) gamma-ray flare observed by \textit{Fermi}-LAT from the blazar 3C 279. It has long been considered a candidate site for accelerating particles to ultra-high energies (UHE) and producing subsequent secondaries. Such an event can be crucial to explore the different phenomena of secondary production from the UHEs and viable to understand the energetics of the sources. Our study finds that the multi-wavelength flare, spanning UV, optical, X-rays, and HE gamma rays, originates from leptonic emissions, whereas the delayed VHE activity by proton synchrotron emission within the source, results from the extended duration of particle acceleration. To explain the prolonged electromagnetic emission,
our model requires a magnetic field luminosity (L$'_B$)\; $6.1 \times 10^{43}$ erg/sec, a proton luminosity (L$'_{p}$)\; $1.2
 \times 10^{46}$ erg/sec in the jet frame.}




\maketitle
\section{Introduction}
\label{introduction}

The observation of TeV-PeV neutrinos from blazar TXS 0506+056 \citep{TXS0506+056} and Seyfert NGC 1068 \citep{ngc1068} has encouraged hadronic-originated emission studies from active galactic nuclei (AGN). The brightest flat-spectrum radio quasar (FSRQ) 3C 279 at a distance z= 0.536 \citep{1965ApJ...142.1667L} can be considered a potential source of hadronic emissions. This source (4FGL J1256.1-0547) shows rapid variability in all wavelengths, including HE $\gamma-$rays, as reported by Compton Gamma Ray Observatory (CGRO) \citep{1992ApJ...385L...1H} and \textit{Fermi} Large Array Telescope (LAT)\footnote{\url{https://fermi.gsfc.nasa.gov/science/instruments/lat.html}} \citep{2015ApJ...803...15P, 2015ApJ...807...79H,2016ApJ...824L..20A}. It is the first quasar observed above 50 GeV by the Major Atmospheric Gamma Imaging Cherenkov Telescopes (MAGIC) in February and March 2006 \citep{2008Sci...320.1752M}. The source was again observed with VHE emissions on 16 January 2007, during a major optical flare, with a flux of $(3.8 \pm 0.8) \times 10^{-11}$ photons $\text{cm}^{-2} \text{s}^{-1}$ above 150 GeV energy \citep{MAGIC3c279}. These observations encourage exploring different HE phenomena study with 3C 279.   
\\
\\
Interestingly, \textit{Fermi}-LAT reported a flaring state of 3C 279 with a peak flux $(8.4\pm0.5) \times 10^{-6}$ photons $\text{cm}^{-2} \text{s}^{-1}$ on 15 January 2018 \citep{Fermi-2018}. The optical follow-up search by Rapid Eye Mount telescope (REM)\footnote{\url{https://www.eso.org/public/teles-instr/lasilla/rem/}} reported magnitudes for optical and near-infrared (NIR) wavelengths, $V=14.39\pm0.05$, $R = 13.96\pm0.14$, $I = 13.37\pm0.08$ $ J = 12.10\pm0.02$, $H = 11.27\pm0.04$, $K=10.27\pm0.04$ \citep{REM-2018}. The Astronomical Observatory of the University of Siena also marked a rise in R-band magnitude \citep{Optical-2018}. {\cite{2019ZAHIR.484.3168S} have analyzed the ultraviolet (UV), optical, and X-ray wavelength events of \textit{Swift} satellite for ten epochs from 17 January 2018 to 1 February 2018 and inferred consistency with REM observed optical data. The analysis also resulted in a possible negligible variation of UV and optical flux for a longer period than the HE-flare observed by \textit{Fermi}-LAT. An additional follow-up survey by Astrorivelatore Gamma a Immagini LEggero (AGILE) reported an increased flux of $(1.8 \pm 0.3) \times 10^{-5}$\; photons $\text{cm}^{-2} \text{s}^{-1}$ above $ 100$\;\text{MeV} \citep{AGILE-2018}. The Dark Matter Particle Explorer (DAMPE) has reported an increase in photon flux by measuring $~(6.45 \pm 3.63) \times 10^{-7} $ photons $\text{cm}^{-2} \text{s}^{-1}$ \citep{DAMPE-2018} after 16 January above 2 GeV. The above multi-wavelength observations of optical, X-ray and $\gamma$-rays confirmed that the source was flaring in January 2018. 
\\
\\
  H.E.S.S. missed monitoring the source most timeline of the \textit{Fermi}-LAT flare and reported the first follow-up VHE events above 100 GeV during 27-28 January with 11$\sigma$ significance \citep{HESS-2018,2019Galax...7...20B}. The preliminary analysis also claims the source showed a flaring state for nearly two days in VHE energies during this epoch \citep{Emery:2019llm}. Hence, this flare in VHE emissions can be considered at least eleven days delayed from the \textit{Fermi}-LAT flare peak. However, any VHE monitoring synchronous to the \textit{Fermi}-flare would have differed or supported the claim. Such delayed VHE activity, non-simultaneous to \textit{Fermi}-LAT flare also possibly occurred for the neutrino point source, blazar TXS 0506+056 \citep{2022ApJ...927..197A,sunanda}.
\\
\\
Dedicated attempts have been made to model this source with a) leptonic single zone, where optical to X-ray emission is explained with lepton synchrotron and HE-VHE emissions with Compton up-scattering of external photons (EC) by accelerated leptons \citep{MAGIC3c279,2019ZAHIR.484.3168S}, b) leptonic two/multi zones \citep{MAGIC3c279, Markus} and c) lepto-hadronic channels, where the HE-VHE photons explained with proton-photon (p$\gamma$) interactions \citep{maria}.
 
Several groups also have explained the 2018 HE-flare of 3C 279 with leptonic model \citep{2020RAJPRICE...890..164P, 2019ZAHIR.484.3168S}. {\cite{2020RAJPRICE...890..164P} studied the broadband variability of the source with emissions from electron synchrotron and EC up-scattering of the broad line region (BLR) and disk region photons, and \cite{2019ZAHIR.484.3168S} modeled with emissions from EC of the external torus region.  \cite{Oberholzer:2019dnz} has explained the delayed VHE activity with the lepto-hadronic synchrotron mirror model \citep{Böttcher_2005}. This model attributes the VHE events originated from $\pi^0$ decay from p$\gamma$ interaction where the photons are the Doppler blue-shifted primary electron synchrotron photons reflected by a nearby mirror cloud. 
\\
\\
To explain this possible delay of $ \sim$11 days between the VHE flare and the \textit{Fermi}-LAT HE flare, we propose a single zone lepto-hadronic model. Our model is based on the fact that the acceleration time of protons is longer than electrons due to their heavier mass. This results in a delay in the proton synchrotron emissions compared to leptonic emission. A similar scenario has been discussed in \citet{German}, supporting the delayed proton synchrotron contribution.  In our model, the multi-wavelength emissions in optical-UV with lepton synchrotron and the X-rays to HE are driven by the up-scattering of external photons in the dusty torus (DT) region. In contrast, the delayed VHE emissions are explained by synchrotron emission of accelerated protons. Additionally, we suggest that the shortened VHE flare compared to the time scale of the proton synchrotron is caused by the escape of protons from the environment. 
\\
\\
The paper is arranged as follows: Section \ref{sec:model}  details the multi-wavelength observations for source 3C 279 collected by various detectors during the relevant period, including the \textit{Fermi}-LAT data analysis during the HE-flare. Section \ref{mod} describes the lepto-hadronic modeling to explain HE and VHE activity. A detailed discussion of our results is presented in Section \ref{result}. 

\section{\label{sec:model} Multi-wavelength Observations of 3C 279}
We collected the multi-wavelength data (UV, optical and X-rays) synchronized with the \textit{Fermi}-LAT flare from 14 January 2018 to 22 January 2018 (58132 to 58140 Modified Julian Days (MJD)) from \cite{2019ZAHIR.484.3168S} and the HE data from \textit{Fermi}-LAT analysis discussed in section \ref{secFermi-LAT}}. The VHE light curve data observed by H.E.S.S. from \cite{Emery:2019llm,Oberholzer:2019dnz}.

\subsection{Fermi-LAT HE light-curve of 3C 279}
\label{secFermi-LAT}
\cite{2019ZAHIR.484.3168S, 2020RAJPRICE...890..164P,Emery:2019llm} have confirmed a flaring state of 3C 279 by analyzing a nearly one-month \textit{Fermi}-LAT light curve (LC) within energy 100 MeV to 300 GeV in January 2018.
The \textit{Fermi}-LAT LC for the time 58130-58144 MJD (12 January 2018 to 26 January 2018) is analyzed in a one-day bin within the energy range 0.1-300 GeV. The data for LC generation within the region of interest (ROI) 10$^{\circ}$ centred at the location of the target source has been collected. The spectral shapes of sources within this region are considered variables. We then performed an unbinned maximum-likelihood analysis
 using the software package Fermitools\footnote{\url{https://github.com/fermi-lat/Fermitools-conda/wiki}} version 2.2.0 and the instrument response function $\textup{P8R2{\_}SOURCE{\_}V6}$. The systematic uncertainties become less effective when the Pass8 analysis is selected and photons above energy 100 MeV are collected. This also reduces possible contamination from non-considered (transient) neighboring sources\footnote{\url{https://fermi.gsfc.nasa.gov/ssc/data/analysis/documentation/Pass8_edisp_usage.html}}. All the parameters of the Galactic diffuse model \enquote{${\textup{{gll}{\_}{iem}{\_}v07.fits}}$} and isotropic component \enquote{$\textup{{iso}{\_}{P8R3}{\_}{SOURCE}{\_}{V3}{\_}{v06}.{txt}}$} are kept free within a radius of 5$^{\circ}$ for this analysis. The events are selected considering Front+Back event type (evtype=3), evclass =128, and with a zenith angle cut of 90$^{\circ}$ which excludes the $\gamma$-ray events contaminated by the bright Earth limb. Events with test statistics (TS)$\, > \, $25 have been considered, as this confirms the subtraction of the background. Figure \ref{fig:fit} shows LC data in the top plot for the TS$ \, > \,$9 in each bin, which considers the detection of the target source. The bottom plot of the same figure illustrates the temporal variation of the spectral index for the power-law fit over the same energy range during the specified period. The variation suggests a hardening of the spectral index during the \textit{Fermi}-LAT flare with the value (2.16$\pm$0.02) whereas during the peak of the VHE flare the spectral index is, (2.06$\pm$0.07). However, the later value shows no significant variation within the systematic errors. This result is consistent with the variation study done in \cite{2019ZAHIR.484.3168S} with log parabola fit.
\\
\\
 The duration of the flare is obtained by analysing its decay and rise time. The \textit{Fermi}-LAT flare's decay and rise time have been calculated by fitting the HE $\gamma$-ray LC with the function given in \cite{2010ApJ...722..520A} 
\begin{equation}
\label{eq:pythagorean}
 F(t)=F_b+F_0 \left[\frac{exp(t_0-t)}{T_r}+\frac{exp(t-t_0)}{T_d}\right]^{-1}  \, ,
\end{equation}
where $F_b$ is the constant base flux, $F_0$ is the photon flux at the peak time $(t_0)$, $T_r$ and $T_d$ are the HE flare's rise and decay times, respectively. The details of the HE $\gamma$-ray LC and the best-fitted values of the parameters are shown in Figure \ref{fig:fit}. The fitting results in the rising, decay, and peak time are $1.61\pm0.18$, $2.14\pm0.30$ days, and $58135.9\pm 0.41$ MJD,  respectively. Generally, the flare period is considered as ${2}(T_d+T_r)$ \citep{2010ApJ...722..520A}. Hence, the reference HE flare has a period of nearly six days.

\begin{figure}
    \centering
    \includegraphics[width=9.4cm]{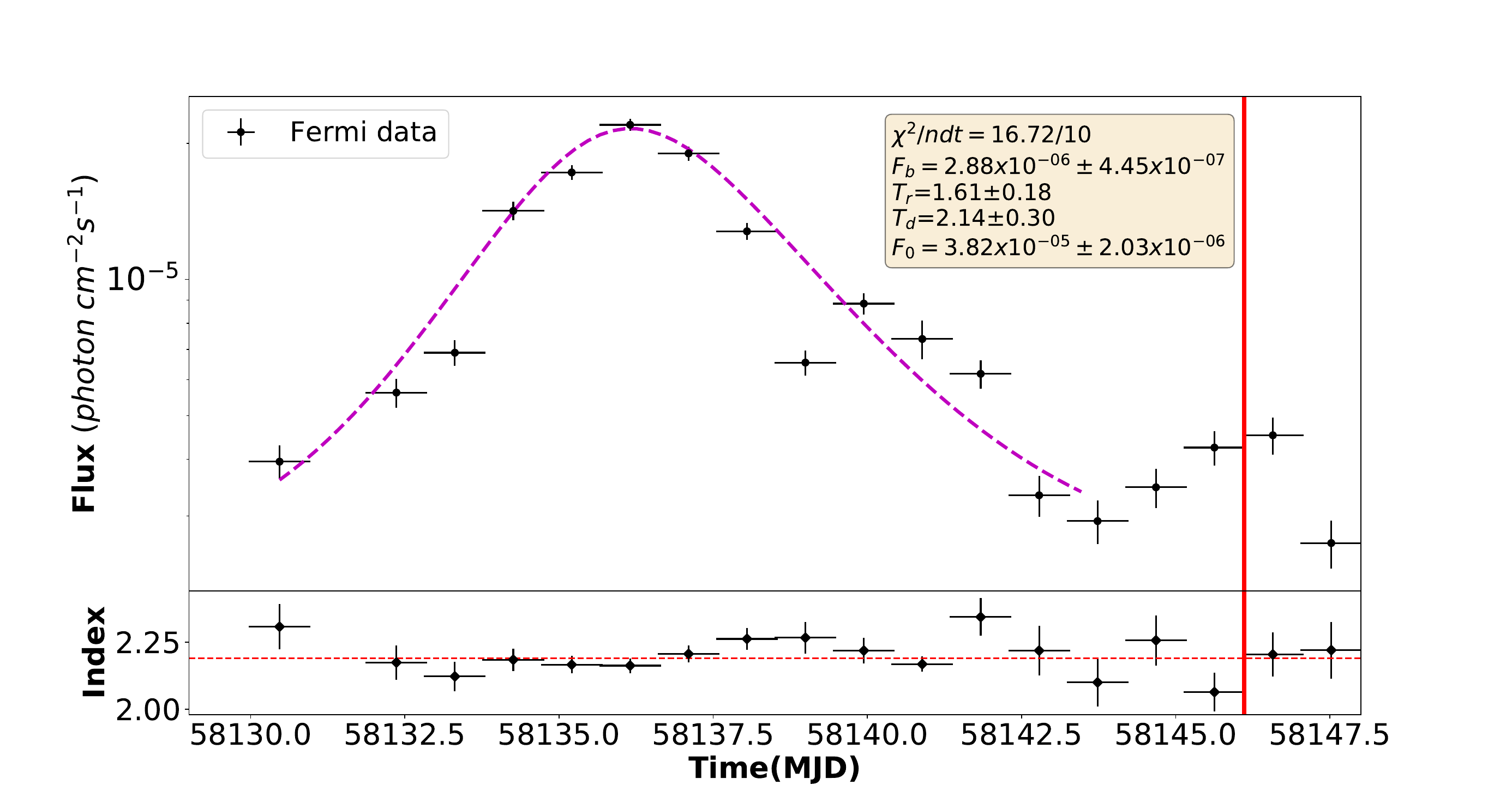}
    \caption{The top plot presents \textit{Fermi}-LAT $\gamma$-ray LC of 3C 279 in the energy range 0.1 to 300 GeV with a one-day bin. The bottom plot illustrates the temporal evolution of the spectral index for a power-law fit.
}
    \label{fig:fit}
\end{figure}

The light curve of HE and VHE activity suggests an almost 6-day HE flare, approximately from 58132 to 58138 MJD with a peak around 58135.9 MJD, and a VHE activity from 58140 to 58147 MJD with a peak flux at 58146.11 MJD \citep{Emery:2019llm} and both LCs are shown in Figure \ref{HE-VHE-lc}. A nearly 11-day delay in VHE activity from the peak of the HE flare can be argued from the LC plot. The (transparent blue) region in Figure \ref{HE-VHE-lc} further confirms the absence of any HE flare during the VHE activity. We propose that this VHE activity is associated with the HE activity through a delayed mechanism. 
\begin{figure}
    \centering
    \includegraphics[width=9.2cm]{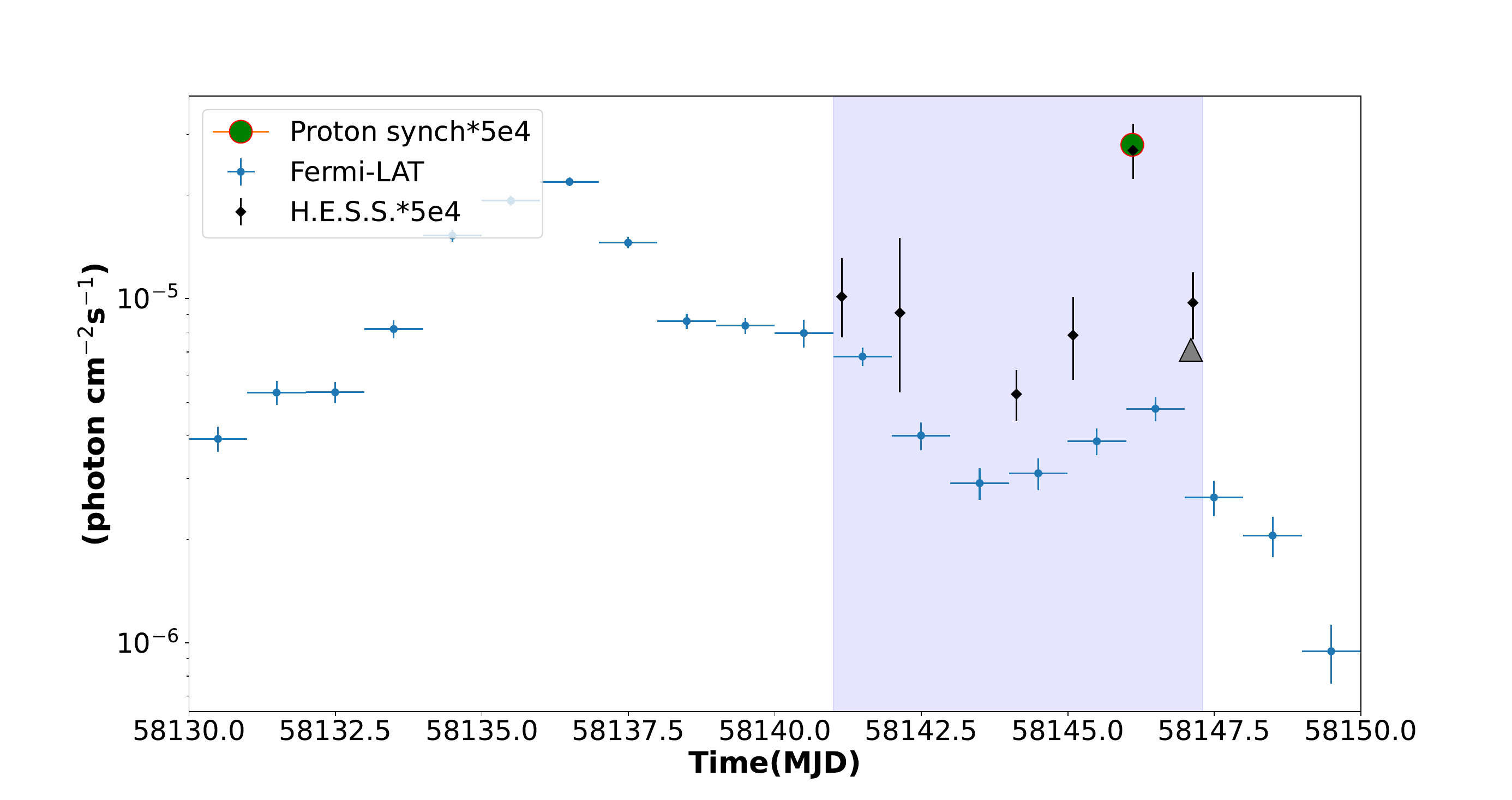}
    \caption{The light curve of \textit{Fermi}-LAT analysis during the period  58130 MJD to 58150 MJD within the energy range 0.1 to 300 GeV. The (blue) circles represent \textit{Fermi} HE-$\gamma$ ray photon flux and the (black) diamond points for VHE activity observed by H.E.S.S. during the period MJD 58140 to MJD 58147, and the (red-green) circle represents the peak (at 58146.11 MJD) energy integrated proton synchrotron emissions from the modeling over the H.E.S.S. energy limit. The (black-Grey) triangle represents the decrease in the integrated synchrotron photon flux after one due to the escape of protons.}
    \label{HE-VHE-lc}
\end{figure}
\subsection{\label{secSED_fit}Spectral Energy Distribution (SED) of the HE flare}
The method described in section \ref{secFermi-LAT} has been followed to collect the SED data for  TS$ \, > \,$9 during this HE flare for the energy range 0.1-300 GeV. The resulting data points and their errors are shown in Figure \ref{fig:my_label}. 

These SED events have been fitted with well-known functions: single power law (PL), broken power law (BLP), and log-parabola (LP) to understand their origin. The functional forms also govern the localization of the source, hence confirming the presence of the target source. The expressions are \footnote{\url{https://fermi.gsfc.nasa.gov/ssc/data/analysis/documentation/Cicerone/Cicerone.pdf}}, 
 
\begin{alignb}
\frac{dN}{dE_{\gamma}}&=  N_0\left(\frac{E_{\gamma}}{E_b}\right)^{-\alpha_1}, \hspace{2.4 cm} \text{PL}\\ \nonumber
       &=N_0\begin{cases}\left (\frac{E_\gamma}{E_b}\right)^{-\alpha_1} \, E_{\gamma}< E_b \\
             \left (\frac{E_\gamma}{E_b}\right)^{-\alpha_2} \, E_{\gamma}> E_b  \;,\hspace{1 cm} \text{BPL}
            \end{cases}       \\ \nonumber
       &=N_0 \left (\frac{E_\gamma}{E_b}\right)^{-\left[\alpha_1+\beta \, ln(E_\gamma/E_b)\right]} .
  \hspace{0.8 cm} \text{LP} 
\end{alignb}
We used Fermitools to fit these functions, with the fitted parameters detailed in Table \ref{tab:2}, the resulting fits are shown in Figure \ref{fig:my_label}.  

To identify the presence of the source at the given position, the TS values for the three functional forms have been calculated. It is defined as, $TS =-2 \,ln( \mathcal{L}_{max,0}/ \mathcal{L}_{max,1})$ \footnote{\url{https://fermi.gsfc.nasa.gov/ssc/data/analysis/documentation/Cicerone/Cicerone_Likelihood/Likelihood_overview.html}}, where $ \mathcal{L}_{max,0}$  denotes the maximum likelihood of the input model corresponding to the null hypothesis, where no additional source is present. $ \mathcal{L}_{max,1}$ is the maximum likelihood of the model with additional source following the functional form at the specified coordinates. A large value of TS indicates the presence of an additional source. The TS values from the \textit{Fermi}-LAT analysis for all three functional forms are given in Table \ref{tab:2}. 

The best functional form representing the production channel of these photons was statistically searched for, according to their Akaike Information Criterion (AIC) value \citep{1974ITAC...19..716A}. AIC is defined as $\text{AIC} =2k-2 \, ln \mathcal{L}$,
where k and $ \mathcal{L}$ denote the free parameters and the likelihood function of the model, respectively. The AIC values for the three models are presented in Table \ref{tab:2}.  The model which has a minimum value of AIC is the preferred model. PL model has the highest AIC value, whereas LP has the lowest value. Hence, LP is the preferred functional form for the SED, consistent with the functional form preference as in \cite{2020RAJPRICE...890..164P}. 
 
\subsection{SED of VHE activity observed by H.E.S.S.}
\label{hess data}
The H.E.S.S. LC above energy 60 GeV of 3C 279 is available in \cite{Emery:2019llm} for 58140 MJD to 58147 MJD and are shown with (black) points in Figure \ref{HE-VHE-lc}. Details of the SED for these VHE events are unavailable, so we calculated a possible value for the spectrum from the LC. We assume the observed VHE photons follow a power-law spectrum in energy ${dN_{\gamma}}/{dE_{\gamma}} \propto (E_{\gamma}/E_{\gamma,0})^{-\alpha_{\gamma}}$. 
Considering the values for the reference energy, $E_{\gamma,0}$ as 200 GeV and the spectral index ${\alpha_{\gamma}}$ as $4.1 \pm 0.68$, we normalized the LC flux at the VHE peak (at 58146.11 MJD) to get the spectrum. The parameter values have been chosen based on the highest VHE activity observed from 3C 279 in January 2006 by the MAGIC telescope \citep{2008Sci...320.1752M}, including the detection of photons with energies up to nearly 500 GeV. The (blue) shaded region in Figure \ref{fig:sed_mod} shows the calculated SED of the VHE activity within the energy range 60 GeV to 500 GeV. The  \textit{Swift}'s Ultraviolet/Optical Telescope (UVOT) events from \cite{2019ZAHIR.484.3168S} corresponding to the HE flare period, from MJD 58134 to MJD 58138, and the period from MJD 58138 to MJD 58142, which is near the timeline of the VHE flare, are also depicted in the same figure using (cyan) circles and (grey) polygons, respectively. 
\begin{figure}
\centering
    \includegraphics[width=9.2cm]{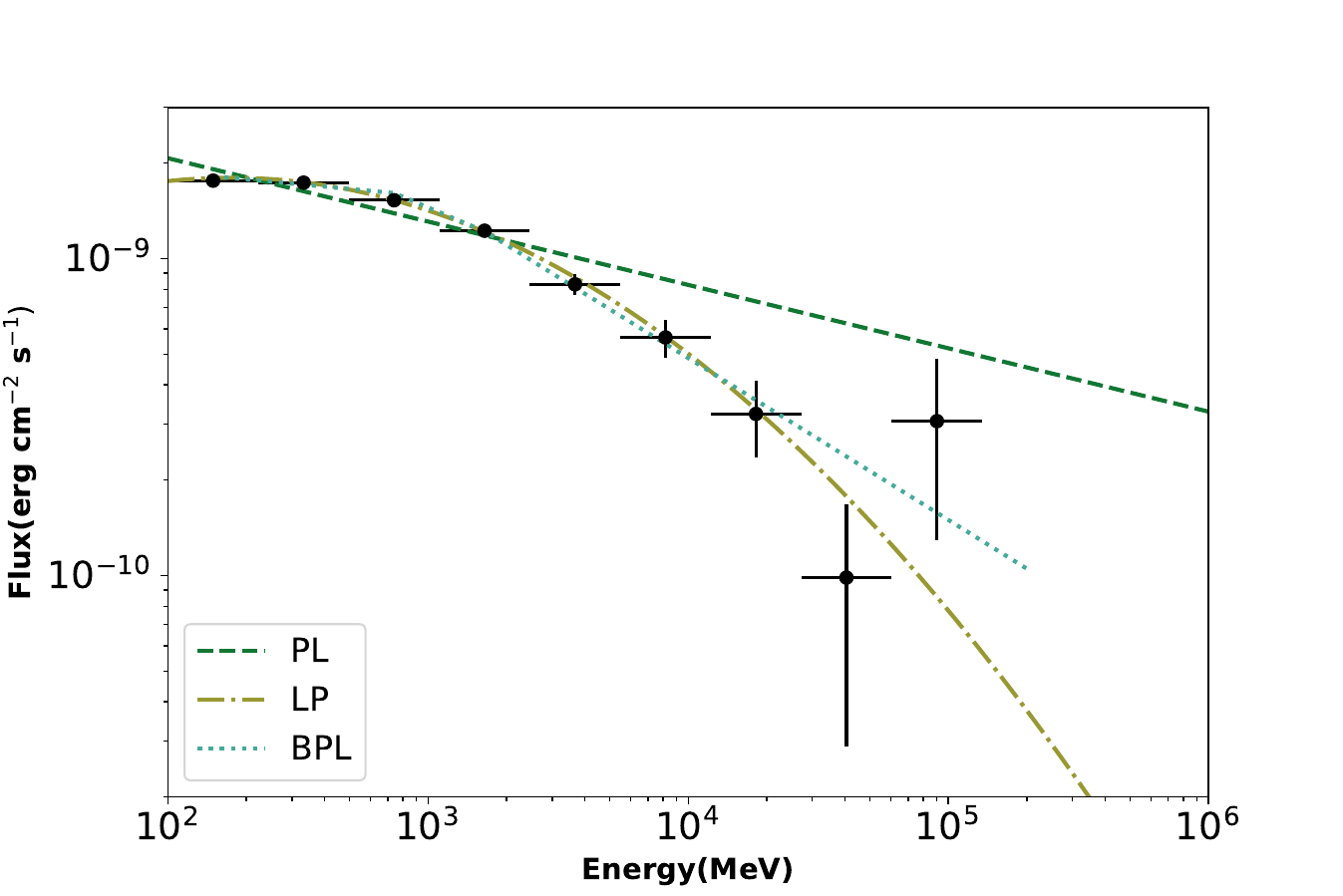}
    \caption{The three fitting functions PL, BPL, and LP to \textit{Fermi}-LAT data and fitting parameters are given in Table \ref{tab:2}. }
    \label{fig:my_label}
\end{figure}
\begin{table}[h]
\begin{tabular}{@{}llll@{}}
\toprule
Model & PL   & BPL & LP\\
\midrule
$\alpha_1$    & $2.20{\pm}0.01$  & $2.07{\pm}0.01$   & $2.14{\pm}0.01$\\
$\alpha_2$    & $--$  & $2.51{\pm}0.04$   & $--$
  \\
$\beta$    & $--$  & $--$   & $0.07{\pm}0.01$  \\
$E_b(\text{MeV})$ & 1000.00 & 1001.32 & 442.00\\
N$_0$($10^{-10}$ TeV$^{-1}$ cm$^{-2}$ s$^{-1}$) & 8.18$\pm$.16 & 9.81$\pm$.01 & 1.01$\pm$.91 \\
AIC & 157820 & 157749.94 & 157749.66 \\
TS & 26475.07 & 26685.22 & 26709.88 \\
\hline
\botrule
\end{tabular}
\caption{\label{tab:2}Fitting parameters of the three functional forms PL, BPL, and LP to the \textit{Fermi}-LAT SED of period 58130-58142 MJD and corresponding AIC and TS values. $\alpha_1, \alpha_2$ and $\beta$ are the spectral indices. $E_b$ and $N_0$ are the break energy and normalization constant, respectively. } 
\end{table}

\section{Lepto-Hadronic Model to explain the HE and VHE activity}
\label{mod}
Our model assumes a spherical, single-emitting blob with radius $R'$, moving along the jet axis within the plasma at a bulk Lorentz factor $\Gamma$. The jet material moves relativistically towards the observer, characterized by the Doppler factor $\delta$. The HE emission originated through the inverse Compton interaction of accelerated leptons with the external DT region photons. However, the VHE emission resulted from the synchrotron emission of accelerated protons in the magnetic field along the path.

\subsection{\label{LMG}Lepton Modeling till HE gamma-rays}

To perform the leptonic modeling of 3C 279 during the observed 6-day flare (58132-58138 MJD, as discussed in section \ref{secFermi-LAT}), we consider a propagated electron spectrum by solving the transport equation of the injected spectrum. The publicly accessible GAMERA code \citep{Hahn} is used to solve the transport equation. The observed multi-wavelength SED events till HE  has been modeled for a total emission period of $\left [6 \, \delta/(1+z)\right]$ days with leptonic emissions. The injected electron spectrum is LP function, $dN/dE_{e}'={N_0}(E_e'/E_{ref}')^{-\left[\alpha_e+{\beta_e}ln(E_e'/E_{ref}')\right]}$, where $E_e' \, \text {and} \, E_{ref}'$ are the electron energy and reference energy, respectively, similar to the functional form of the HE-SED (see section} \ref{secSED_fit}). The primed ($'$) parameters are in the jet frame and parameters without prime are in the observer frame. GAMERA considers the minimum electrons (of mass $m_e$) Lorentz boost $\gamma'_{min}(= E_{e,min}'/(m_ec^2))$, maximum electrons Lorentz boost $\gamma'_{max}(= E_{e,max}'/(m_ec^2))$, spectral indices
$(\alpha_e,\beta_e)$, and the number of electrons per unit energy$(N_0)$ as free parameters. 

 \begin{figure*}[h!]
\includegraphics[width=\textwidth,height=11cm]{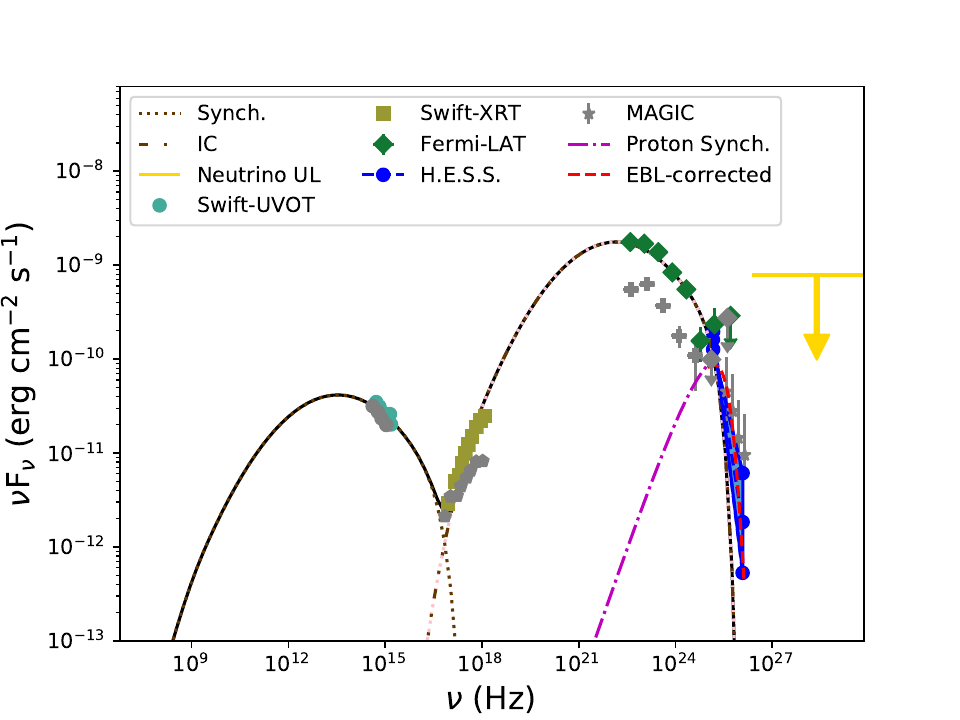}
    \caption{Multi-wavelength Modeling of 3C 279 quasar. The (green) diamond shows the \textit{Fermi}-LAT SED between MJD 58132 and MJD 58138. The (grey) plus sign represents the HE flux at the time (MJD 58142 to MJD 58148) of the VHE flare.   
    The (cyan) circle and (lemon) square points represent the UV, optical, and X-ray for time MJD 58134 to MJD 58138, respectively. The \textit{Swift}-UVOT and XRT for MJD 58138 to MJD 58142,
near the VHE flare timeline, are shown with the  (grey) circle and (grey) polygons, respectively. The (blue) shaded region shows the VHE-flare H.E.S.S. observation.  The VHE-flare SED has been calculated from the light curve in \cite{Emery:2019llm,Oberholzer:2019dnz} considering the spectral index observed by MAGIC in 2006 \cite{2008Sci...320.1752M}  see Section \ref{hess data}. The MAGIC observed events in 2006 are shown with (grey) star points. The electron synchrotron emission is shown as a (brown) dotted line, while a (brown) dash-dot-dotted line represents IC scattering from the DT. A (black) solid line depicts the combined contribution from both interactions. 
The (magenta) dashed dot line and (red) dashed line represent the proton synchrotron flux contribution and correction due to interaction with the EBL, respectively. }
    \label{fig:sed_mod}
\end{figure*}

To model the observed multi-wavelength SED with the above-mentioned propagated electrons, the required values of the parameters are  B$'$= 1.6 G (similar to \cite{2023MNRAS.526.6364T} ) R$'=8.5 \times 10^{16}$ cm, $\Gamma =17.5$ and $\delta=35$, which is above the minimum Doppler factor required for the escape of $\gamma$-rays around 500 GeV \citep{Gammarrayloud}. The emitting region can be considered at a location from the black hole $R_{em}=c\Gamma^2 t_{var}/(1+z)  \sim 9.2 \times 10^{17}$ \textup{cm} for a variability $t_{var}=1.79$ days obtained following \cite{2020RAJPRICE...890..164P} for one-day bin light curve as shown in Figure \ref{fig:fit}.   The Infra-red (IR) torus region can be assumed at a distance $R_{IR}=2.5 \times 10^{18} \, \text{cm} \,L^{1/2}_{d,45} \sim 3.53 \times 10^{18} \;\text{cm}$ as reported in \cite{2009MNRAS.397..985G, Pian_1999} from black hole, with the disk (UV) luminosity $L_{d} =10^{45}\;L_{d,45}\;\text{erg/sec}$ \citep{Pian_1999}. The photon density of the DT region can be expressed as U$'_{DT}=\left[\Gamma^2 L_{d} \zeta /4 \pi c R_{IR}^2\right]$ erg/cm${^3}$ following \cite{1997ApJ...484..108S} where $\zeta=0.3$, resulting $U'_{DT}$=0.039 erg/cm$^3$ and the temperature of the DT region is T$'_{DT}=1000$ K.
The lepton modeling of the SED for the multi-wavelength observations is shown in Figure \ref{fig:sed_mod} and the modeling parameters are listed in Table \ref{tab:parmod}.

\subsection{Hadron modeling for VHE gamma-rays activity}

\begin{figure}
    \centering
       \includegraphics[width=9cm]{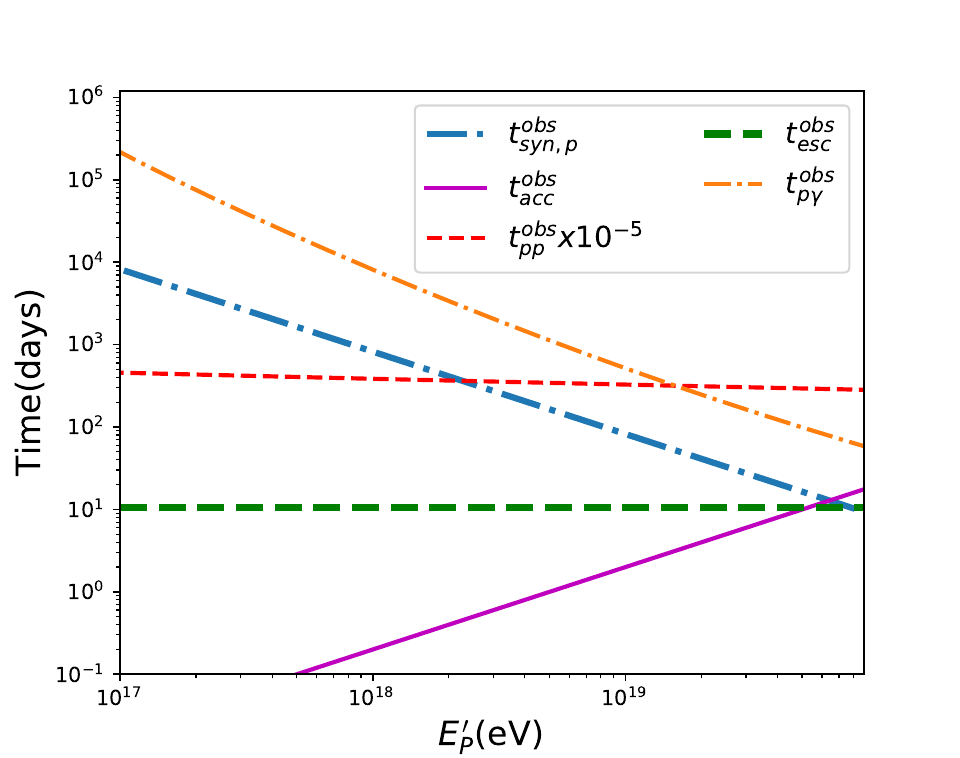}
    \caption{Time scale of the proton cooling processes in the observer frame. The (magenta) solid line represents $t_{acc}^{obs}$, the three cooling channels $t_{pp}^{obs}$, $t_{p\gamma}^{obs}$ and $t_{syn,p}^{obs}$ are presented with the (red) dashed line, thin (orange) dashed-dot line and thick (blue) dash-dot line, respectively. The (green) solid dashed line represents the observed escape time $t_{esc}^{obs}$.} 
    \label{fig:mod_time}
\end{figure}

We model the approximate 11-day delayed VHE activity observed by H.E.S.S., with protons synchrotron due to the long acceleration period of protons. The protons acceleration time is $t^{obs}_{acc}\simeq{1.4}\;{E'_{p,19}} \;{\delta^{-1}_{35}}\;{B'^{-1}_{1.6} \; {\eta_{5}(E'_p)}}\;(1+z)\; \textup{days}$ following \cite{2000NewA....5..377A}, and protons synchrotron cooling timescale  is
$
t^{obs}_{syn,p}=58\;B_{1.6}'^{-2}\;E'^{-1}_{p,19}\;\delta_{35}^{-1}\;(1+z)\; $\textup{days} in the observer frame,  where $E'_{p}=10^{19}\;E'_{p,19} \;\text{eV}$, $\delta=35\;\delta_{35}$, \;$B'=1.6\;B'_{1.6}\;\text{G}$ and gyro factor $\eta(E'_p)$ is in units $\eta_{acc}(E'_p)=5\;\eta_5(E'_p)$ . The maximum energy achieved by accelerated electrons is much less than the accelerated protons due to their cooling in the source's environment. In Figure \ref{fig:mod_time}, the (magenta) solid  line represents the protons acceleration time, while the  (blue) thick  dash-dot line represents the synchrotron time, using the fitted parameters of the Lepton model.

The other dominant channel for cooling of protons is through p$\gamma$ interaction. The accelerated protons interact with the external DT photons and electron synchrotron emissions. The cooling time scale of p$\gamma$ interaction with a photon density of  $n(\epsilon')$ at energy $\epsilon'$ is 
$t_{p\gamma}' \simeq\; 0.03\;c\; \langle \sigma_{p\gamma} f \;\rangle\; n(\epsilon')\ E_{p,19}'^{-1}\;\text{sec}$
suggested in \cite{Aharonian2002}, where $\langle \sigma_{p\gamma}f \rangle\simeq 10^{-28}\;\text{cm$^2$}$ is the photo-meson cross-section.  The variation of the p$\gamma$ production timescale with protons energy in the observer frame is depicted by a (orange) thin  dash-dot line in Figure \ref{fig:mod_time}. It is important to note that this process is significantly suppressed compared to synchrotron losses. The other possible cooling process of the accelerated protons is by interacting with the ambient protons. The cooling timescale for this channel is, $t_{pp} = (K\sigma_{pp}n_p)^{-1}\;\text{sec}$. Here, we have taken the ambient density ($n_p$) as $3\times 10^6 $ \; cm$^{-3}$ nearly close to the value in \cite{banik}, and inelasticity $K=0.17$, as interpreted in \cite{PhysRevD.74.034018}, $\sigma_{pp}$ is the pp interaction cross-section \cite{PhysRevD.79.039901}. The same figure shows the pp channel timescale with a (red) dashed  line normalized with $10^{-5}$. It suppresses in most of the accelerated-protons energy ranges. 
\\
\\
The diffusion of particles in the emission region decreases the number of accelerated particles, lowering the emission flux. In the case of protons synchrotron emission as a model, different escape time scenarios have been proposed according to different energy-dependent diffusion regimes as suggested in \cite{Basumallick}. We assume an energy-independent escape time, $t^{obs}_{esc} = \eta_{esc}\;R/c\;(1+z)/\delta\; \text{sec}$, where $\eta_{esc} = 8$. A (green) solid  dashed line represents this in Figure \ref{fig:mod_time}
for protons, using the multi-wavelength modeling parameters.

The same figure shows the protons can be accelerated till several factors of $10^{19}$ eV. The escape time dominates over all other cooling processes of the protons at nearly all energies.  The resulting escape fraction of protons can be calculated as $f_{esc}=(1-t'_{esc}/t'_{c})$ as described in \cite {PhysRevD.90.023007}, where $t'_{c}$ is cooling time. Hence, the corresponding survival protons in the emission region have a fraction $(1-f_{esc})$ of the injected protons.

\subsection{ VHE emissions with Protons Synchrotron} 

We considered the spectrum of the injected accelerated protons as log parabola distribution,  $N(\gamma'_p)=k\;\gamma_p^{-\alpha_l}$, here $\alpha_l=\alpha_p+\beta{\ln{\gamma'}}$ where $\alpha_p$, $\beta_p$ are spectral indices and $k$ is the normalization constant. The synchrotron emissions of the above protons spectra for energy-independent escape time within the protons energy range in the magnetic field $B'$  is calculated following 
\cite{2002MNRAS.332..215A,sunanda} to explain the VHE activity events. 
The protons at energy $E'_p$ is related to the critical frequency of synchrotron photons ($\nu_c$ in the observer frame) as  
\begin{equation}
E'_p=2.17 \times 10^{7} \textup{eV} \, \left(\frac{\nu_{c}{(1+z)}}{B'\;\delta}\right)^{1/2} 
\label{epvsnuc} ,
\end{equation}
from \cite{2000NewA....5..377A}.
  The relativistic protons are accelerated up to a maximum energy ${E_{p, max}^{\prime}}$ in the blob, where the maximum energy is adopted from Figure \ref{fig:mod_time} following $t_{esc}=t_{acc}$ \citep{PhysRevD.90.023007}. The Hillas criterion allows the protons to accelerate till energy, $E'_{p,max}= 4.41 \times 10^{19} B'_{1.6} R'_{16}$, $\text{eV}$. Proton energy $(E'_p)$ of $ 3.9 \times 10^{19}$ $\text{eV}$ is required to explain the emission of a critical energy photon of 500 GeV in the observer frame as per Equation  \ref{epvsnuc}.
The synchrotron photon radiation at frequency $\nu'$ generated for the protons spectrum \citep{2011hea..book.....L} is 
\begin{equation}
\label{flux}
    J'(\nu')d\nu' =-m_pc^2\frac{d\gamma'_p}{dt'}N(\gamma'_p)d\gamma'_p   \;\;.
\end{equation}

We calculated the protons energy loss rate, $-{d\gamma'_p}/{dt'} ={4}/{3}\;\sigma_T\; c\; U'_{mag}\; {\gamma'_p}^2 {m_e^2}/{m_p^2}$.
\;The observed synchrotron photons flux then can be calculated as,
 \begin{alignb}
F(\nu)\;d\nu&=  \frac{\delta^2\;V'\;(1+z)^2}{4\pi\;d_L^2}\;J'\left[\frac{(1+z)\nu}{\delta}\right]\;d\nu     \\
       &= A\;\nu^{-b} \;d\nu  ,
\label{prot-syn-flux}       
\end{alignb}

where $V'$ and $d_L$ are the emission region volume and the luminosity distance,  respectively. The constant A =$k\;(m_pc^2)^{(1-\alpha_l)}/(48\;\pi^2d^2)\;(\sigma_T\;m_e^2c/m_p^2)\;(e/2\pi\;m_pc)^{(\alpha_l-3)/2}\\\times(\delta/(1+z))^{(\alpha_l+5)/2}\;(B^{(\alpha_l+1)/2})$  and $b =(\alpha_l{-}$1)/2. 

The contribution of protons synchrotron flux obtained from Equation \ref{prot-syn-flux} is shown in Figure \ref{fig:sed_mod} with the (magenta) dashed-dot line. The parameters for the hadronic model are listed in Table \ref{tab:parmod}.
\begin{table}[h!]
	\centering
	\caption{SEDs modelling parameters for 3C 279}
	\label{tab:parmod}
	\begin{tabular}{lccr} 
	   \hline
		Parameters& Values \\
            \hline
            z & $0.536$ \\
        d(Mpc) & $2.29\times10^3$\\
        \hline
            $\delta$ & $35$\\
            B(G) & $1.6$ \\
                   $R' $ (cm)& $8.5\times10^{16}$ \\
\hline
            $\alpha_{e}$ & $2.0$\\
            $\beta_e$  & $0.44$\\
            ${\gamma_{e,min}'}$ & ${1}$\\
            
            ${\gamma_{e,max}'}$& ${1.5\times10^4}$\\
             
   \hline     
         $\alpha_p$ &{2.0}\\
         $\beta_p$ &{0.0012}\\
      $\gamma_{p,min}'$&$1$\\
        $\gamma_{p,max}'$& $4.1\times10^{10}$\\
    \hline
    \hline
             $L'_e({erg}\;{s^{-1}})$ & $3.6\times10^{41}$\\
             $L'_B({erg}\;{s^{-1}})$ & $6.1\times10^{43}$\\
        $L'_p({erg}\;{s^{-1}})$ & $2.6\times10^{45}$\\
        \hline
        $T'_{DT}$(K)  & ${1000}$ \\
        $U'_{DT}(\text{erg/cm$^3$})$ & $0.039$\\

   \hline
	\end{tabular}
\end{table}
\section{Results \& Discussion}
\label{result}
The \textit{Fermi}-LAT light curve of the quasar 3C 279 in the period 58130 to 58142 MJD within the energy range 0.1 to 300 GeV suggests a flaring state in HE. However, the sub-TeV gamma-rays observed with the H.E.S.S. telescope indicate VHE activity with a nearly 11-day delay from the peak of the HE flare. We explain both these events with a lepto-hadronic model where the observations of optical and UV are explained with electron synchrotron emissions and the HE-flare with the electron-EC by the DT region photons.  The SED modeling insists the lepton luminosity of $L'_e =1.4\times10^{41}$ erg/sec using  {$L_e' ={\pi}\;{R'^{2}_b}{c}\int_{\gamma'_{e,min}}^{\gamma'_{e,max}} m_e{c}^2{\gamma'_e}{N(\gamma')} \,d\gamma'$} from \cite{baniklumino} in a magnetic field of 1.6 G. 

We model the 11-day delayed VHE activity with protons synchrotron emissions in the same magnetic field B$'=$1.6 G. The delay is explained by the protons' acceleration time, $t_{acc}^{obs}$. Over $\Delta t= 11$ days, the position of the emission region can be anticipated to shift to a distance $ \sim c \delta \Delta t/(1+z)= 1.5\times 10^{18}\;\textup{cm}$.  The jet with pure toroidal magnetic field will result in the variation of the magnetic field over a distance ($l$) from the black hole as $B'\propto l^{-1}$ \citep{Blandford}. Due to the travel of the emitting region, a change of magnetic field $\sim 0.25$ can be expected. Subsequently, a change will be observed in the flux of electron synchrotron emission as $B'^{(\alpha_e+1)/2}$. Hence, the decrease in synchrotron flux after 11 days is expected to be 0.14. This value will decrease further if considered for a pure poloidal magnetic field in the Jet. However, the \textit{Swift}-UVOT and optical data analysis from \cite{2020RAJPRICE...890..164P,2019ZAHIR.484.3168S} suggests insignificant variation in flux over the period MJD 58132 to 58145 within statistical error bars. Consequently, we assume the magnetic field of the Jet environment remains constant over this prolonged time, and the protons continue to accelerate and radiate through synchrotron emission in this magnetic field to explain the VHE events. Additionally, the electrons of the blob continue to produce synchrotron emissions in the regime of the optical-UV band. However, the movement of the blob possibly changed the environment from a higher density of external photons to a lower density of photons, which lowered the HE flux.

To explain the VHE-SED observed by H.E.S.S., we considered an energy-independent escape time. The acceleration time explains the 11$-$day delay in the protons synchrotron. Our model suggests $\eta_{esc}=8$ for escape time and $\eta_{acc}=5$, for acceleration time both depend on the microscopic physics of the jet, a field that requires further investigation. We explicitly show the protons synchrotron emission radiation in the magnetic field of 1.6 G in Figure \ref{fig:sed_mod} with a (magenta) dash-dot line. We also show the VHE  $\gamma$-ray emissions from the source after correction due to interaction with the Extragalactic background light (EBL)\footnote{\url{http://www.astro.unipd.it/background/}} with the dashed (red) line in the same figure for the source at a distance $d= 2.29 \times 10^9$ pc. Our model requires the total proton luminosity to be {$L'_{p} ={\pi}\;{R'^{2}_b}\;{c}\int_{\gamma'_{p,min}}^{\gamma'_{p,max}} m_p{c}^2{\gamma'_p}{N(\gamma')} \,d\gamma'$= $1.2\times 10^{46}$ \;erg/sec  with magnetic luminosity $L'_{B}=\pi\;{R'^{2}_b}\;{c}\;U'_B=6.1\times 10^{43}$ erg/sec, where $U'_B$ is magnetic field density.
Again, for a black hole of mass, $M_{BH} =8\times10^8\textup{M}_\odot$ (\cite{2009A&A...505..601N}), the Eddington luminosity for the source is $1.04\times10^{47}\textup{erg/sec}$. According to \cite{2019NatAs...3...88G,2014Natur.515..376G,2001APh....15..121M} the required injected proton luminosity is  $L^{in}_p \simeq \Gamma^{2}L'_p$, is $7.9\times10^{47}\;\textup{erg/sec}$. This value indicates that $L^{in}_p$ requires super-Eddington luminosity to explain the sub-TeV emissions. Previous studies also suggest high super-eddington luminosity for hadronic channels \citep{PhysRevD.107.103019,2013ApJ...768...54B}.\\

Our model suggests that 64$\%$ of the protons remain in the blob, based on $(1-f_{esc})$, after accounting for the escape time. We calculated the integrated synchrotron-radiated proton flux over the energy range from 60 GeV to 500 GeV. The resulting flux at the peak of the VHE activity, on 58146.11 MJD, is shown in the LC see Figure \ref{HE-VHE-lc} with a (red-green) circle. The VHE activity is expected to last for a period similar to $t_{acc}^{obs}$, but the observed activity only lasts for about two days, as seen in the same figure. We propose that the shorter flare duration is due to the escape of a fraction of the protons, assuming that the same fraction survives into the following day. This decrease in proton number, corresponding to the photon number, can be verified using the LC in Figure \ref{HE-VHE-lc}. The resulting integrated photon flux for the next day is shown in the same figure with a (black-grey) triangle.
\\
\\
We target one such observed event that reports nearly an 11$-$day delay in the VHE event from one of the extraordinary quasars, 3C 279. We explain this delay originated from a protons synchrotron channel compared to the historical model of electron IC. More such multi-wavelength follow-up observations would help significantly identify the origin of VHE activities in blazars. A detailed generic model with different physical parameters for the hadronic channel is also required to understand further such events.

\bmhead{Acknowledgements}

We thank Foteini Oikonomou for her valuable suggestions, which improved the manuscript significantly. We would also like to acknowledge Abhradeep Roy and Pratik Majumdar for their valuable discussions on \textit{Fermi}-LAT data analysis. Additionally, Sunanda extends gratitude to Aaqib Manzoor for insightful discussions on modeling. Sunanda would like to acknowledge the CSIR for providing a fellowship and fruitful discussion with Soebur Razzaque.


\bibliography{reference}

\end{document}